\documentclass[twocolumn,showpacs,preprintnumbers,amsmath,amssymb]{revtex4}

\usepackage{graphicx}
\usepackage{dcolumn}
\usepackage{bm}

\newcommand{\bra}{\langle}
\newcommand{\ket}{\rangle}

\begin{document}


\title{Nonlinear Elasticity of Single Collapsed Polyelectrolytes}

\author{Hirofumi Wada}
\email{wada@daisy.phys.s.u-tokyo.ac.jp}
\affiliation{%
Department of Physics, University of Tokyo, Hongo, Tokyo, 113-0033,
Japan}%
\author{Yoshihiro Murayama}
\affiliation{%
Department of Physics, University of Tokyo, Hongo, Tokyo, 113-0033,
Japan}%

\author{Masaki Sano}
\affiliation{%
Department of Physics, University of Tokyo, Hongo, Tokyo, 113-0033,
Japan}%

\date{\today}

\begin{abstract}
Nonlinear elastic responses of short and stiff 
polyelectrolytes are investigated by dynamic simulations on a 
single molecule level.
When a polyelectrolyte condensate undergoes a mechanical unfolding,
two types of force-extension curves, i.e., a force plateau
and a stick-release pattern, are observed depending on the 
strength of the electrostatic interaction.
We provide a physical interpretation of such force-extension
behavior in terms of intramolecular structures of the condensates.
We also describe a charge distribution of condensed counterions onto 
a highly stretched polyelectrolyte, which clarifies a formation of
one-dimensional strongly correlated liquid at large Coulomb 
coupling regime where a stick-release pattern is observed.
These findings may provide significant insights into the relationship
between a molecular elasticity and a molecular mechanism of 
like-charge attractions observed in a wide range of charged 
biopolymer systems.
\end{abstract}

\maketitle
Electrostatic interactions in aqueous media can be controlled by
changing temperature, dielectric constant of the solvent, and
counterion valency~\cite{israelachvili,fuoss}. 
In a strongly charged system, correlations between counterion
charge fluctuations that are neglected within the Poisson-Boltzmann (PB) theory 
give rise to various dramatic phenomena which apparently contradict our 
understandings based on the PB theory~\cite{oosawa,rouzina,shklovskii,netz,arenzon}.
One of the most familiar example is probably the condensation of a single 
DNA molecule in the presence of multivalent cations~\cite{widom,bloomfield,stevens}, 
which indicates the counter-intuitive ``like-charge attractions''~\cite{gensen-ha}.
A variety of problems related with electrostatic effects is being of
great interest in soft condensed matter physics today~\cite{gelbart}.
Ionic effects also 
modifies elastic properties of PE chains drastically~\cite{skolnick,
odijk,stevens2,barrat,orland,baumann2}.
Recent micromanipulation experiments have shown impressive elastic 
responses of single collapsed DNA molecules~\cite{baumann,murayama}, 
which significantly deviate from those expected from the standard 
worm-like chain (WLC) model~\cite{marko}.
In spite of such experimental evidences, theoretical and numerical
studies of this problem have been lacked so far, mainly because of its
nonequilibrium nature combined with long-ranged Coulomb 
interactions.

In this Letter, we report on results of the Brownian Dynamics
(BD) simulations on the stretching of single PE condensates.
The present simulation covers only a salt-free system with a  
short chain strand, 
but effects of counterions which are strongly coupled to PEs are 
taken into account explicitly.
Force-extension ($f$-$x$) curve of a PE chain
shows the standard WLC elasticity, a force plateau, and a 
stick-release pattern as the electrostatic coupling increases.
This trend is consistent with that observed in the DNA stretching
experiments~\cite{baumann,murayama} where the electrostatic coupling 
strength is controlled by the concentration of the added multivalent 
cations.
The structure factor of condensed couterions onto a stretched
PE reveals a formation of one-dimensional (1D) strongly 
correlated liquid (SCL)~\cite{rouzina,shklovskii,netz,arenzon}
when a $f$-$x$ curve shows a stick-release, 
while such a charge ordering is absent in the case of a force plateau.
In order to highlight electrostatic effects on the 
macroscopic polymer elasticity, we are primarily focusing on systems 
with very strong Coulomb interactions
which is beyond a typical physiological condition.
Furthermore the present study deals only with relatively short and stiff polymers
and also neglects complicated structures of biopolymers~\cite{allahyarov}, direct 
comparison with any experiment is therefore impossible at this stage.
The primary aim in this Letter is to build bridges between the macroscopic
molecular elasticity and the theoretically predicted SCL picture,
which may serve as a first step towards microscopic understandings
of available experiments.

\begin{figure*}
\includegraphics[width=17cm]{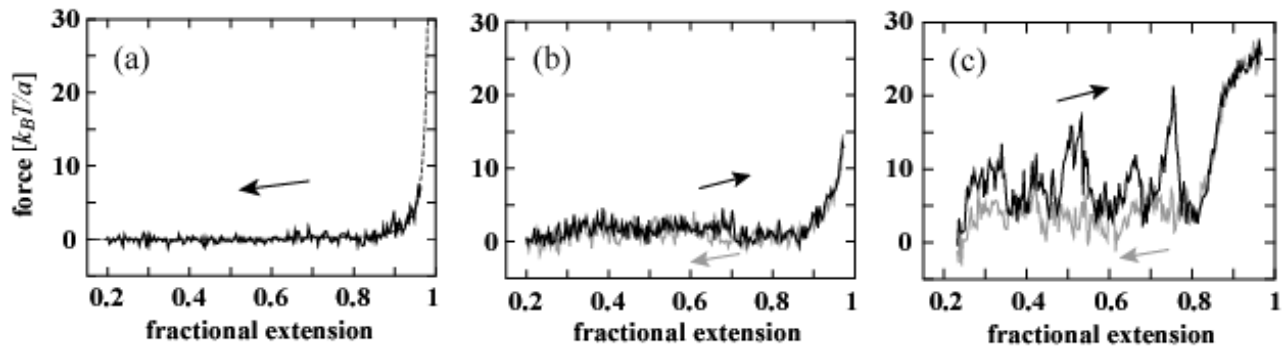}
\caption{\label{fig:fxcurve} The force-extension curves
of a PE for three different values of the coupling parameter 
(a) $\Gamma=0.1$ (WLC), (b) $\Gamma=40$ (force plateau), 
and (c) $\Gamma=110$ (stick-release). 
The dashed line in (a) is the best fit of Eq. (\ref{wlc}) to
the simulation data. Arrows indicate stretching/relaxing.}
\end{figure*}

In our BD simulations, the chain comprises $N$ harmonically 
linked beads of radius
$a$, each of which carries a charge $-qe$. The system also includes
$N$ oppositely charged counterions of the same valency and radius
to keep the overall charge neutrality.
The charged particles interact via the unscreened Coulomb 
interaction, where the solvent is replaced by a background 
with the dielectric constant $\varepsilon$.
The system is placed in a cubic box and periodic
minimum-image boundary conditions are imposed to keep 
the monomer/counterion density 
$\rho_{mon}=\rho_{ion}\approx 1.5\times 10^{-5}/a^3$ constant
\cite{winkler,netz2}.
Long time dynamics of monomers and counterions are
well described by the position Langevin equations~\cite{doi} 
\begin{eqnarray}
 \gamma\frac{d{\bf r}_i}{dt} &=& 
	-\nabla_{{\bf r}_i}U[{\bf r}(t)]+{\bf f}_i(t),
 \label{eq01}
\end{eqnarray} 
where $\gamma$ is the damping constant, $U$ the potential energy 
and ${\bf f}_i$ the vectorial random force acting on 
the particle $i$, which satisfies the correlation property
$\bra{\bf f}_i(t){\bf f}_j(t')\ket=
2\gamma k_BT \delta_{ij}{\bf 1}\delta(t-t').$
Because we are primarily concerned with dense ionic globules, 
hydrodynamic interactions are not taken into account 
in this study~\cite{winkler,netz2}.
The particle radius $a$, the energy scale of the Lennard-Jones
potential $\epsilon$ 
and $\tau=\gamma a^2/\epsilon$ are chosen as 
the unit of length, energy, and time in our simulations.
The potential energy is composed of several
kinds of contributions: $U=U_s+U_b+U_c+U_{LJ}$.
The first one $U_s$ ensures the connectivity of a PE.
To maintain small variances of bond lengths around $b$, i.e.,
the chain length around $L=b(N-1)$,
we use a large value of the spring constant $K=400\epsilon/a^2$.
The second one $U_b$ accounts for the {\it intrinsic} bending 
stiffness of a chain:
\begin{eqnarray}
 U_b &=& \kappa\sum_{i=1}^{N-2}\left(1-\frac{
	({\bf r}_{i+1}-{\bf r}_i)\cdot ({\bf r}_{i}-{\bf r}_{i-1})}{b^2}
		\right).
 \label{pot02}
\end{eqnarray}
The intrinsic persistent length is thus 
given by $L_p^0=b\kappa/k_BT$.
The Coulomb potential $U_c$ acts between all charged particles 
except between bonded ones. 
In order to prevent a collapse of monomers and counterions,
a truncated Lennard-Jones potential $U_{LJ}$ acting
between all particles is also incorporated
only for $|{\bf r}_i-{\bf r}_j|<b$.
The coupling constant $\Gamma=q^2l_B/a$ controls the relative
strength of the electrostatic interaction to the thermal energy
at a distance $a$.
(The Bjerrum length $l_B=e^2/4\pi\varepsilon k_BT \approx 0.7$ nm 
in water at room temperature.)

In the beginning of our simulation, a randomly generated 
initial configuration is allowed to collapse to a globule 
and equilibrated for $5\times10^6$ time steps 
with $\Delta t=10^{-3}\tau$.
Then the collapsed chain is stretched by pulling its one end at a 
constant speed, while the other end is fixed in a position.
The stretching speed $v_s$ is measured by the value $\mu$ defined 
as $\mu=v_s/v_0$, where $v_0=b/\tau_R$ with 
$\tau_R=\gamma b^2N^2/3\pi^2\epsilon$ the Rouse relaxation time for
a flexible polymer with $N$ monomers at $k_BT=\epsilon$ \cite{doi}.
In all simulations presented below, we fix $N=32, b=2a, L/L_p^0=4, 
\mu=0.1$ and $k_BT=0.3\epsilon$, while $\Gamma$ is widely changed.
To complete single stretch/relax cycle, simulation runs for more than
10$^8$ time steps even for $N=32$.
Simulations for larger $N$ is desirable, but it is much more time 
consuming because the total time steps is proportional to
$L\tau_R \propto N^3$ in addition to the $O(N^2)$ compulations 
for Coulomb interactions at every time step.

Figure.~\ref{fig:fxcurve} (a) shows the $f$-$x$ curve of the PE
for $\Gamma=0.1$.
The chain remains elongated in this case, and the $f$-$x$ 
curve obeys the standard WLC formula:
\begin{eqnarray} 
 \frac{L_pf}{k_BT} &=& u+\frac{1}{4(1-u)^2}-\frac{1}{4},
 \label{wlc}
\end{eqnarray}
where $u=x/\langle L\rangle$ is the fractional extension~\cite{marko}.
Even in this case,
the persistence length $L_p$ differs from $L_p^0$ because of the 
electrostatic contribution~\cite{skolnick,
odijk,stevens2,barrat,orland,baumann2}.
\begin{figure}
\includegraphics[width=7cm]{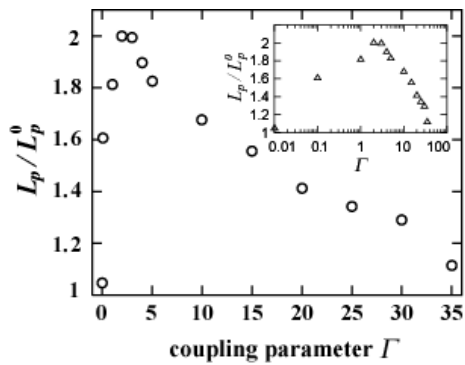}
\caption{\label{fig:persist} The effective persistence length of a PE
deduced from its force-extension curve as a function of $\Gamma$. 
Inset is the same plot on a semi-log scale.}
\end{figure}
In Fig.~\ref{fig:persist}, we plot the effective $L_p$ 
as a function of $\Gamma$, which are determined by fitting
Eq.~(\ref{wlc}) to the simulated $f$-$x$ curves for $0.80\leq u\leq 0.97$.
For small $\Gamma$ (roughly less than 3), 
the electrostatic repulsion between monomers leads to a 
stretched configuration of the chain, i.e., an increasing stiffness.
As the coupling increases, however, it is more and more screened
out by the counterion condensation.
Far from the onset value of the Manning condensation 
($\Gamma_c\approx 2$ for infinitely long, stiff chain~\cite{manning}), 
the electrostatic interactions are significantly reduced, 
and $L_p$ again decreases to $L_p^0$.
Such behavior is also consistent with that of the radius of gyration
$R_g$ observed in other simulations~\cite{winkler,netz2}, if one recalls
that the persistent length is related to the radius of gyration
as $R_g^2\sim L_pL$ for a semiflexible polymer.
The sudden drop of $L_p$ at $\Gamma=35$ may be the sign of intramolecular
collapse~\cite{golestanian}.

For $\Gamma=40$ [Fig.~\ref{fig:fxcurve} (b)], 
a distinct force plateau is emerged in the $f$-$x$ curve, 
reflecting intramolecular attractions which are purely 
electrostatic origin.
To evaluate the magnitude of the condensation force between monomers,
we calculate the excess work $\Delta W=\int_{x_1}^{x_2}
(f-f_0)dx=\bar{f}(x_2-x_1)$ from the data, where $f_0$ is the WLC force 
calculated by Eq. (\ref{wlc}),
and $x_1$ and $x_2$ are chosen as $0.3\langle L\rangle$ and 
$0.8\langle L\rangle$.
We obtain $\Delta W \sim 0.8 k_BTL/a$, which roughly 
corresponds to $\bar{f}\sim 1.6 k_BT/a$.
Taking into account that $\bar{f}$ is a certain increasing function
of $\Gamma$, let us tentatively compare our $\bar{f}/\Gamma$ 
with that of available experiments.
Assuming DNA's linear charge density $\nu=q/(2a)$ and the 
charge valency $q=4$ in our model, we have $a\sim 0.34$ nm
(1bp), which gives $\bar{f}/\Gamma \sim 0.04 k_BT$/bp for $\Gamma=40$. 
On the other hand, for DNA with trivalent cation $(q=3)$ 
at room temperature, one finds $\Gamma=2\nu l_B/a \cong 24$. 
Using the experimental value $\Delta W=0.33k_BT$/bp for 
trivalent cation CoHex obtained by Baumann {\it et al.} \cite{baumann},
one would obtain $\bar{f}/\Gamma\sim 0.01 k_BT$/bp for $\Gamma=24$,
which agrees with our value.

\begin{figure}
\includegraphics[width=8.0cm]{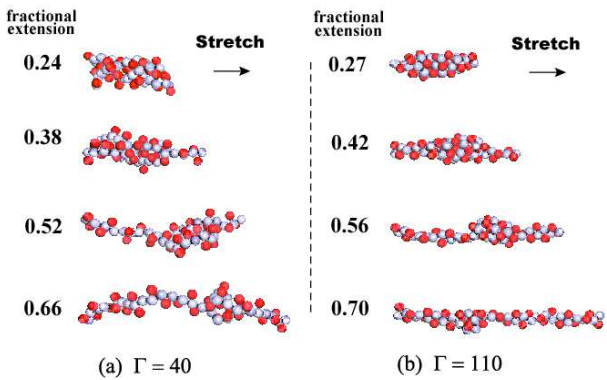}
\caption{\label{fig:spshots} Snapshots of a PE during stretching
for (a) $\Gamma=40$ and for (b) $\Gamma=110$.
Corresponding $f$-$x$ curves are Fig. \ref{fig:fxcurve} (b)
and (c). 
Dark and light spheres represent counterions and monomers,
respectively. For these coupling strengths, all the counterions are 
condensed onto the PE.}    
\end{figure}

Several snapshots of the PE during stretching
are displayed in Fig.~\ref{fig:spshots} (a).
When the tension is applied externally, the ionic condensate
(the PE-counterions complex) can change its overall structure 
to a more energetically stable one, because intramolecular
attractions are not so strong (typically a few $k_BT$ as 
inferred above) for the present moderate $\Gamma$.
Therefore, the PE can change its configuration continuously to
minimise the stored elastic energy during 
stretching/relaxing, which we consider leads to  
the appearance of the force plateau.

In Fig.~\ref{fig:fxcurve} (c), we show the $f$-$x$ curve for the 
largest coupling parameter: $\Gamma=110$.
The curve exhibits a pronounced sawtooth pattern, which may 
reflect intermittent plastic deformations of the glass-like ionic 
condensate~\cite{winkler}. 
(As will be described below, the condensed counterions are still in 
a liquid phase, but their individual motions are extremely slow).
Figure~\ref{fig:spshots} (b) shows a series of the corresponding snapshots of 
the PE, which clearly demonstrates the coil-globule coexistence
within the single PE under the imposed extension~\cite{cooke}.
This is a noticeable difference with the case in an external electric
field where an unfolding of a PE is rather abrupt~\cite{netz2}.
For sufficiently large $\Gamma$, the electrostatic attractions are so 
strong that the PE condensate cannot alter its structure easily.
Then the elastic energy is more and more stored in the chain
as the extension increases, until the PE condensate is locally unravelled
in rather abrupt way.
\begin{figure}
\includegraphics[width=5.3cm]{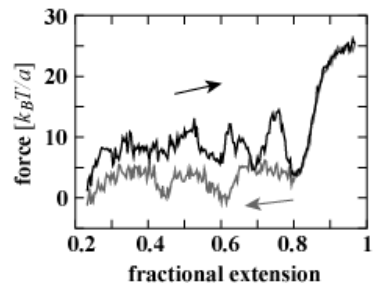}
\caption{\label{fig:fxavg}Force-extension curve
averaged over five independent runs starting from the same initial 
collapsed state for $\Gamma=110$.}
\end{figure}

It is important to note that there is marked hysteresis
in the $f$-$x$ curve [Fig.~\ref{fig:fxcurve} (c)] during stretching 
and relaxing.
This suggests that folding and unfolding 
transitions of the complex are first-order-like, namely, 
they do not take place at the same extension~\cite{cooke}.
This feature is more distinctive in the $f$-$x$ curve
[Fig.~\ref{fig:fxavg}] averaged
over several independent runs with the same initial
collapsed state for $\Gamma=110$. 
This averaged curve is smoothed out in comparison to a single-pass 
sawtooth pattern as Fig.~\ref{fig:fxcurve} (c), 
which again indicates that each stick-release response is caused by a
nonequilibrium configurational change of a PE condensate.
Averaging over a large number of single-pass $f$-$x$ curves,
one would obtain a smooth $f$-$x$ curve with vanishing peaks.
However a hysteresis would remain even in that limit, as long as 
a time scale of $v_s$ is faster 
than the longest transition time of a PE condensate to overcome an 
intramolecular local energy barrier~\cite{wada}. 

\begin{figure}
\includegraphics[height=5.0cm]{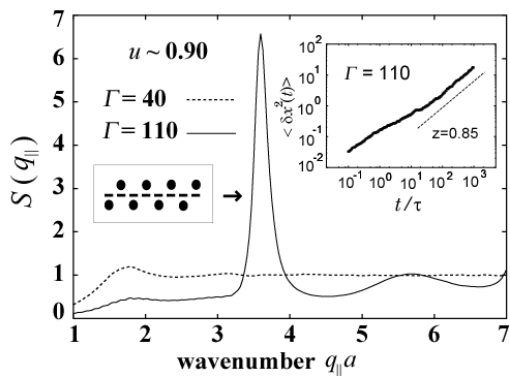}
\caption{\label{fig:sq} 1D structure factor $S(q_{\parallel})$ of 
counterions condensed onto a stretched PE with its
fractional extension $u$ kept nearly $0.9$ 
for $\Gamma=40$ (broken line) and for $\Gamma=110$ (solid line).
Inset: The self-diffusion of a single condensed counterion 
for $\Gamma=110$, which exhibits sub-diffusive behavior given by 
$\langle\delta x^2(t)\rangle \propto t^z$ with $z=0.85$
for $t/\tau>10^1$.}
\end{figure}
We now turn our attention to the counterion charge distribution 
along a highly stretched PE chain.
In fig.~\ref{fig:sq}, 1D structure factor
$S(q_{\parallel})$ is shown for $\Gamma=40$ and $110$, with the PE
fractional extension $u$ kept nearly $0.9$.
Here $S(q_{\parallel})$ is calculated according to 
$S(q_{\parallel})=\frac{1}{N}\sum_{i,j}
\langle \exp[iq_{\parallel}(x_i(t)-x_j(t))]\rangle$, 
where $x_i(t)$ is the $x$ coordinate of the $i$-th 
counterion and the bracket 
represents long-time average (taken as roughly $10^7$ time 
steps).
One can find a crystalline order of condensed counterion for 
$\Gamma=110$, while a gaslike phase is found for $\Gamma=40$.
The most prominent peak at $q_{\parallel}a \sim \pi$ for $\Gamma=110$
is identified as the Coulombnic ground state configuration of the
condensed counterion.
However, a self-diffusion of a counterion
(averaged over all particle motions) is sub-diffusive,
though extremely slow, even for $\Gamma=110$ (see the inset of 
Fig.~\ref{fig:sq}).  
The condensed counterions therefore form a 1D SCL at the surface of a PE.
We now expect as follows.
A long-ranged attraction mediated by long-ranged charge fluctuations~\cite{gensen-ha,
golestanian} gives rise to the force
plateau in $f$-$x$ curve; on the other hand, the highly nonequilibrium 
elasticity such as the stick-release for large $\Gamma$ mainly comes from 
the short-ranged strong attraction created by the structural 
correlations of 1D SCL. 
This attraction is predicted to dominate over
the long-ranged one at a very short distance that is typically within
the inter-particle distance $b$ \cite{lau}.
Because the counterions are much more strongly bound to the PE
for $\Gamma=110$ than for $\Gamma=40$ (see fig.~\ref{fig:spshots}), 
it is reasonable to consider that different mechanisms of 
a net attraction between similarly charged monomers are
underlying for $\Gamma=40$ and 110.
To verify this idea, however, fine data for
larger $N$ systems are required because we have to diminish finite-size effects
and influences of system-size dependent duration time of metastable states
as much as possible.

We expect that a sufficiently large $\Gamma$ regime is 
theoretically accessible from the first principle using the 
recently developed strong-coupling theory~\cite{netz,naji}, 
while for intermediate $\Gamma$ regime, a phenomenological 
approach which postulates a ``short-ranged'' intramolecular 
attraction can be still useful to understand a macroscopic 
elasticity of a long PE~\cite{wada}.
(Note that the intramolecular attraction for moderate $\Gamma$ is 
long-ranged, as mentioned, compared with a length scale $a$, 
but is still significantly ``short-ranged'' compared with an overall 
polymer contour length which often extends up to several $\mu$m
for biopolymers such as DNA.)

In summary, we have studied force responses of collapsed short PE strands
by the dynamic simulations in a wide range of $\Gamma$.
The simulation results are consistent with the experimental trends,
and the idea is proposed by analysing the simulation data 
that the SCL state of the counterion may be responsible for 
the stick-release response. 
To unambigiously conclude these findings, more work for
larger $N$ systems will be required.

We thank N. Yoshinaga for useful discussions.
This work was supported by Grant-in-Aid for JSPS
Fellows for Young Scientists, from Ministry of Education,
Culture, Sports, Science, and Technology, Japan.


\begin{thebibliography}{99}
\bibitem{footnote} Experimentally, the effective charge valency of couterions 
nearby a PE, which controlls the strength of the electrostatic coupling 
parameter, increases as the concentration of added multivalent cations
is increased because of the entropically driven counterion exchange
between mono- and multi-valent cations.
\end{thebibliography}
\end{document}